\documentclass[showpacs,amssymb,aps]{revtex4}
\newcommand{\be}{\begin{equation}}
\newcommand{\ee}{\end{equation}}

\newcommand{\del}[1]{\delta_{#1,0}}

\newcommand{\nnbb}{\nonumber\\}

\newcommand{\tr}{{\rm Tr\,}}
\newcommand{\un}{${\rm U}(N)\;$}
\newcommand{\sun}{${\rm SU}(N)\;$}
\newcommand{\pop}{{\sigma}}
\newcommand{\tpt}{{\tau}}
\usepackage{epsf}
\usepackage{graphics}

\begin{document}

\title{Hard Thermal Effects in Noncommutative \un Yang-Mills Theory}

\author{F. T. Brandt$^a$, J. Frenkel$^a$ and
D. G. C. McKeon$^{a,b}$}
\affiliation{$^a$ Instituto de F\'{\i}sica,
Universidade de S\~ao Paulo,
S\~ao Paulo, SP 05315-970, BRAZIL}
\affiliation{$^b$ Department of Applied Mathematics, The University of
Western Ontario, London, ON  N6A5B7, CANADA}


\begin{abstract}
We study the behaviour of the two- and three-point thermal Green
functions, to one loop order in noncommutative \un Yang-Mills theory, 
at temperatures $T$ much higher than the external momenta $p$.
We evaluate the amplitudes for small and large values of the variable 
$\theta\,p\,T$ ($\theta$ is the noncommutative parameter) and exactly 
compute the static gluon self-energy for all values of
$\theta\,p\,T$. We show that these gluon functions, which have a
leading $T^2$ behaviour, are gauge independent and obey simple Ward
identities. We argue that these properties, together with the results 
for the lowest order amplitudes, may be sufficient to fix uniquely the 
hard thermal loop effective action of the noncommutative theory.
\end{abstract}


\maketitle

\section{Introduction}\label{sec1}
Noncommutative manifolds have been used in physics for quite some 
time \cite{Groenewold:1946}
and were more recently applied in the context of string theories
\cite{Connes:1998crSeiberg:1999vs}. In certain circumstances, the
low-energy behaviour of these theories may be described in terms of
gauge fields defined on a space-time where the coordinates do not
commute, so that 
\be
\left[x_\mu,x_\nu\right] = i\, \theta_{\mu\nu}.
\label{1}\ee
The antisymmetric tensor $\theta_{\mu\nu}$, which has the canonical dimension
of inverse mass squared, is assumed to be independent of the
space-time coordinates. 
Such gauge field theories are non-local and exhibit
many intriguing properties, which have been much studied in recent years
(for reviews and a complete list of references see,
for example, \cite{Douglas:2001ba,Szabo:2001kg}. In particular,
several thermal effects in noncommutative theories have been already
examined in a series of interesting
papers \cite{Arcioni:1999hw,Fischler:2000fv,Fischler:2000bp,Landsteiner:2000bwlandsteiner:2001ky}.

In gauge theories at finite temperature, a consistent perturbative
expansion requires the resummation of a set of diagrams called 
{\it hard thermal loops} \cite{braaten:1990azBraaten:1990kk}. These
arise from one-loop diagrams in the region where the internal momenta
are of order of the temperature $T$, which is large compared with all
the external momenta. The purpose of this work is to study in the
non-commutative \un Yang-Mills theory, the high temperature behaviour
of the two- and three-point gluon functions. Our method of calculation
employs an analytic continuation of the imaginary-time 
formalism \cite{kapusta:book89lebellac:book96das:book97}. Using this
approach, we relate the Green functions to forward scattering
amplitudes of on-shell thermal particles, a technique that has been
previously applied in the \sun gauge theory as well as in gravity 
\cite{frenkel:1991ts,brandt:1993dk,brandt:1997se}. In contrast to the
situation in the commutative theory, where the hard thermal loop
contributions are completely determined in the region where 
$T\gg p$, one has to consider in the noncommutative case also the value of
the independent parameter $\theta\, p \, T$. For arbitrary values of
this parameter, the explicit calculation of these non-local amplitudes
is, in general, very difficult. Only in certain limiting cases, such
as $\theta\, p\, T\ll 1$ or $\theta\, p\, T\gg 1$, is that their
evaluation becomes more transparent. An exception occurs in the static
case, where one can evaluate, as shown in section \ref{sec2}, 
the gluon self-energy $\Pi_{\mu\nu}^{AB}$ in a closed form for all
values of $\theta\, p \, T$. The corresponding result shows a
significant difference from the behaviour in the commutative case,
where only the $\Pi_{00}^{AB}$ component survives in the static
limit. On the other hand, in the noncommutative theory, we find that
also the $\Pi_{ij}^{AB}$ components receives leading $T^2$
contributions in the static case. This behaviour reflects the effect of
extra magnetic fields which are induced by the noncommutative
character of the theory. Furthermore, we show in the section
\ref{sec2} that the gluon self-energy is generally transverse with
respect to the external momenta, and that all the $\theta$-dependence
resides in the ${\rm U}(1)$ subgroup of \un. 

The three-point gluon function is discussed in section \ref{sec3},
where we also provide its full $\theta$-dependence which arises in
both the ${\rm U}(1)$ and \sun sectors. Although the inherent angular
integrations are extremely involved and cannot be performed in closed
form, we demonstrate that, in general, the leading $T^2$ contributions
to the three-point amplitude are related to those of the two-point
function in a manner dictated by the Ward identity.

Our main motivation for this investigation is that, under certain
conditions, these gauge invariant amplitudes, together with the Ward
identity, may be sufficient to determine the effective action which
sums up the effects of all hard thermal loops. We discuss this issue
in section \ref{sec4} and present more details on our calculations in
the appendices.

\section{The two-point function}\label{sec2}
\begin{figure}[h!]
\includegraphics*{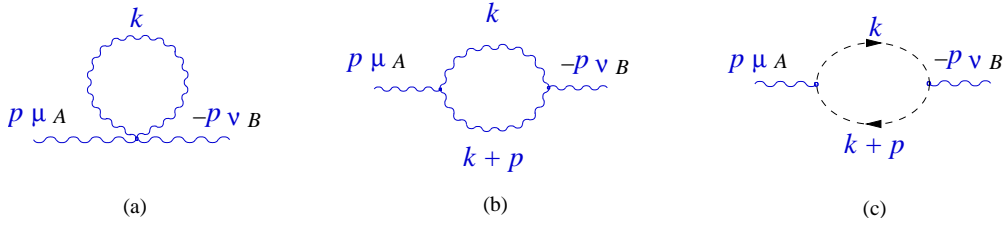}
\caption{One-loop diagrams which contribute to the self-energy
in the noncommutative \un theory. 
Wavy and dashed lines denote respectively gauge particles and ghosts.
The external momenta are inward.} \label{fig1}
  \end{figure}

\begin{figure}[h!]
\includegraphics*{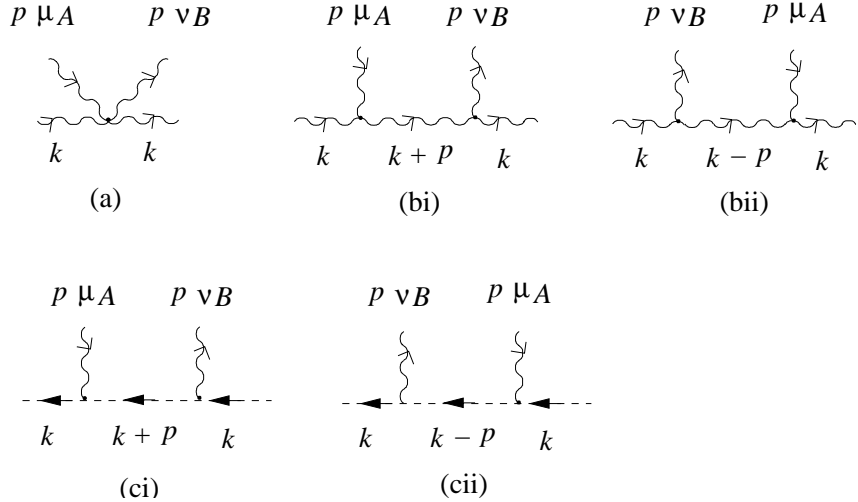}
\caption{Forward scattering amplitudes corresponding to the
diagrams in Figure \ref{fig1}. The direction of the ghost momentum 
is to the left (the same as the corresponding internal gluon line). 
Contributions with
$k\rightarrow -k$ are to be understood.} \label{fig2}
  \end{figure}

In accordance with the approach initiated in \cite{barton:1990fk} and
extended to the Yang-Mills theory in \cite{frenkel:1991ts} we can compute 
the Feynman graphs of Fig. (\ref{fig1}) by considering 
the on-shell forward 
scattering amplitudes ${\cal A}_{\mu\nu}^{AB}$ of Fig. (\ref{fig2}). 
These amplitudes are related to the corresponding two-point function
by the equation
\be
\Pi_{\mu\nu}^{AB} = -{1\over (2\pi)^3}\int{d^3\vec k\over 2\,|\vec k|} N(|\vec k|)
\left.{\cal A}_{\mu\nu}^{AB}\right|_{k_0=|\vec k|};\;\;\;\; A,B =
0,1,2, \cdots,N^2-1,
\label{4}\ee
where
\be
N(|\vec k|) = {1\over {\rm e}^{|\vec k|/T} -1}
\ee
is the Bose-Einstein distribution function. 

Using the Feynman rules shown in the appendix \ref{appA}, 
the contribution of the graph of Fig. (\ref{fig2} a) to the 
forward scattering amplitude is given by 
\be
{B_{AB}^{\mu\nu}}_{(a)} = -3\, g^2\eta^{\mu\nu}
\left[\tr D_A\,D_B\left(1-c_p\right) - \tr F_A\,F_B\left(1+c_p\right)\right].
\label{6}\ee
We are employing the definitions 
\be
{(F_A)}_{BC} \equiv f_{BAC},\;\;\; {(D_A)}_{BC} \equiv d_{BAC},
\label{7}\ee
as well as the abbreviations
\be
c_p\equiv\cos(p\times k),\;\; c_{pq}\equiv\cos(p\times q),\;\;
c_{{1\over 2}p}\equiv \cos({1\over 2}p\times k),\;\;
s_{{1\over 2}p}\equiv \sin({1\over 2}p\times k),\;\;
c_{{1\over 2}p\times q}\equiv \cos({1\over 2}p\times q);\;\;\;
p\times q\equiv p_\mu \theta^{\mu\nu} q_\nu
\label{8}\ee
Using the results (see \cite{Armoni:2000xrBonora:2000ga} for
similar formulas in the context of one-loop renormalization 
of noncommutative theories)
\begin{eqnarray}
\tr F_A\, F_B & = &  -N(1-\del{A})\,\delta_{AB} \nnbb
\tr D_A\, D_B & = &\;\;N(1+\del{A})\,\delta_{AB} \nnbb
\tr D_A\, F_B & = & 0,
\label{9}\end{eqnarray}
Eq. (\ref{6}) reduces to
\be
{B_{AB}^{\mu\nu}}_{(a)} = - 6\, g^2\,N\,\delta_{AB}\eta^{\mu\nu}
\left(1-\del{A}\,c_p\right).
\label{10}\ee

From the graph of Fig. (\ref{fig2} bi) we receive a total contribution
\begin{eqnarray}
{B_{AB}^{\mu\nu}}_{(bi)} & = & {g^2\over 4} \, N\, \delta_{AB}
\left[5p^2\eta^{\mu\nu} -2p^\mu p^\nu+5\left(k^\mu p^\nu+k^\nu p^\mu\right)
+10 k^\mu k^\nu+2k\cdot p\eta^{\mu\nu}\right]
{1\over p^2 + 2\,k\cdot p} \nnbb
& \times &\left[D_A\,D_B(1-c_p) - F_A\,F_B(1+c_p)\right] + 
(k\rightarrow -k).
\label{11}\end{eqnarray}
The contribution of Fig. (\ref{fig2} bii) is identical to that of 
Fig. (\ref{fig2} bi) with the momentum reversed. Consequently,
totalling Figs. (\ref{fig2} bi) and (\ref{fig2} bii) results in
\be
{B_{AB}^{\mu\nu}}_{(b)} =  {g^2} \, N\, \delta_{AB}
\left(1 - \del{A}\,c_p\right)
{5p^2\eta^{\mu\nu} -2p^\mu p^\nu+5\left(k^\mu p^\nu+k^\nu p^\mu\right)
+10 k^\mu k^\nu+2k\cdot p\eta^{\mu\nu}
\over p^2 + 2\,k\cdot p} + (k\rightarrow -k).
\label{12}\ee
In a completely analogous fashion, we find that the 
amplitude
receives the following contributions from Figs. (\ref{fig2} ci) and
(\ref{fig2} cii),
\be
{B_{AB}^{\mu\nu}}_{(ci)}  =  -{g^2} \, N\, \delta_{AB}
\left(1 - \del{A}\,c_p\right)
{k^\mu \left(k+p\right)^\nu\over p^2 + 2\,k\cdot p} + (k\rightarrow -k).
\label{13}\ee
and
\be
{B_{AB}^{\mu\nu}}_{(cii)}  =  -{g^2} \, N\, \delta_{AB}
\left(1 - \del{A}\,c_p\right)
{k^\mu \left(k-p\right)^\nu\over p^2 - 2\,k\cdot p} + (k\rightarrow -k).
\label{14}\ee

In computing (\ref{10}), (\ref{12}), (\ref{13}) and (\ref{14}) the
gauge parameter $\xi$ in Eq. (\ref{apa1}) has been taken to be
arbitrary, but it cancels completely in the final result for the $T^2$ terms.     
In the regime in which $p\ll k\sim T$, we make the expansion
\be
{1\over p^2 + 2\,k\cdot p} = {1\over 2\,k\cdot p} -
{p^2\over (2\,k\cdot p)^2} + \cdots,
\label{15}\ee
so that at the {\it leading order} in $(2\,k\cdot p)^{-1}$, the total
contribution to ${\cal A}^{AB}_{\mu\nu}$ coming from (\ref{10}),
(\ref{12}), (\ref{13}) and (\ref{14}) is
\be
{\cal A}_{\mu\nu}^{AB} =-4 g^2 N \,\delta^{A,B}\, \left[1-\delta^{A,0}
\cos(\tilde p\cdot k)\right] G_{\mu\nu};\;\;\; 
\tilde p_\mu\equiv p^\nu\theta_{\nu\mu}
\label{16}\ee
where
\be
G_{\mu\nu} = {p^2 k_\mu k_\nu\over (k\cdot p)^2} -
{p_\mu k_\nu + p_\nu k_\mu\over k\cdot p} + \eta_{\mu\nu}.
\label{17}\ee

One can easily verify that the transversality property
\be
p^\mu \Pi_{\mu\nu}^{AB} = 0
\label{18}\ee
is satisfied. Indeed, this is a direct consequence of 
$p^\mu G_{\mu\nu}=0$. Therefore, we can express the self-energy in
terms of the following decomposition
\begin{eqnarray}
\Pi_{\mu\nu}^{AB} & = & 
  \Pi_1^{AB}\left(\eta_{\mu\nu}-{p_\mu p_\nu\over p^2}\right) 
+ \Pi_2^{AB}\left(p_\mu - {p^2\over p\cdot u}u_\mu\right)
       \left(p_\nu - {p^2\over p\cdot u}u_\nu\right){1\over p^2}
\nonumber\\
& + & \Pi_3^{AB}\, p^2 \tilde p_\mu \tilde p_\nu +
\Pi_4^{AB}\left[\left(p_\mu - {p^2\over p\cdot u}u_\mu\right)\tilde p_\nu +
           \left(p_\nu - {p^2\over p\cdot u}u_\nu\right)\tilde p_\mu\right],
\label{19}\end{eqnarray}
where $u_\mu$ represents the heat bath four velocity [$u=(1,0,0,0)$].
A straightforward calculation gives
\be
\Pi_1^{AB} = \Pi^{\mu\;AB}_{\;\;\mu} + (\pop^2-1) \Pi_{00}^{AB} - 
{\tilde p^\mu \tilde p^\nu\over \tilde p^2} \Pi_{\mu\nu}^{AB};\;\;\;
\pop^2\equiv {p_0^2\over \vec p^2}
\label{20}\ee
\be
\Pi_2^{AB} = \pop^2\,\Pi^{\mu\;AB}_{\;\;\mu} + 
2\pop^2 (\pop^2-1)\Pi_{00}^{AB} 
- \pop^2 {\tilde p^\mu \tilde p^\nu\over \tilde p^2} \Pi_{\mu\nu}^{AB}
\label{21}\ee
\be
\Pi_3^{AB} = -{1\over \tilde p^2 p^2}\Pi^{\mu\;AB}_{\;\;\mu} -
{\pop^2-1\over \tilde p^2 p^2} \Pi_{00}^{AB} + 
2 {\tilde p^\mu \tilde p^\nu\over \tilde p^4 p^2} \Pi_{\mu\nu}^{AB}
\label{22}\ee
\be
\Pi_4^{AB} = (\pop^2-1) {p_0 \tilde p^\mu \over \tilde p^2 p^2} \Pi_{\mu 0}^{AB}.
\label{23}\ee
In order to ensure unitarity \cite{Douglas:2001ba,Szabo:2001kg}, we
have taken
\be
\theta_{0\mu} = 0.
\label{24}\ee
The computation of the integrals appearing in Eqs. (\ref{20}) to
(\ref{23}) is carried out in the appendix \ref{appB}. We find 
\be
\tilde p^\mu \Pi_{\mu 0}^{AB} = 0,
\label{25}\ee
\be
\Pi^{\mu\;AB}_{\;\;\mu}  = N\,\delta^{A,B}\,{g^2\,T^2\over 3} 
\left[1 -  {6\over \pi^2} \delta^{A,0}
\sum_{n=1}^{\infty}{1\over n^2 + \tpt^2}\right],
\label{26}\ee
\begin{eqnarray}
\Pi^{AB}_{00} & = & {2\,g^2N\,\delta^{A,B}\,\,T^2\over(2\pi)^2}
\int_{-1}^{1} d\zeta
\left[1 -{2\pop\over \pop - \zeta} + 
{\pop^2-1\over(\pop - \zeta)^2}\right]
\nnbb &\times &
\left[{\pi^2\over 6}-
\delta^{A,0}\,\sum_{n=1}^\infty 
{n\over [n^2+\tpt^2(1-\zeta^2)]^{3/2}}\right] .
\label{27}\end{eqnarray}
and
\begin{eqnarray}
{\tilde p^\mu \tilde p^\nu\over \tilde p^2} \Pi_{\mu\nu}^{AB}
& = & -{2\,g^2N\,\delta^{A,B}\,\,T^2\over(2\pi)^2}
\int_{-1}^{1} d\zeta \, \left\{
\left[{\pop^2-1\over(\pop - \zeta)^2}{1-\zeta^2\over 2}-1\right]\right.   
\nnbb & \times &   \left[{\pi^2\over 6}-\delta^{A,0}\,\sum_{n=1}^\infty 
{n\over [n^2+\tpt^2(1-\zeta^2)]^{3/2}}\right]
\nnbb & + & \left.
\delta^{A,0}
{\pop^2-1\over(\pop - \zeta)^2}{1\over 2 \tpt^2}
\sum_{n=1}^\infty 
{\left(n-\sqrt{n^2+\tpt^2(1-\zeta^2)}\right)^2
\left(n+2\sqrt{n^2+\tpt^2(1-\zeta^2)}\right)
\over \left(n^2+\tpt^2(1-\zeta^2)\right)^{3/2}}
\right\},
\label{28}\end{eqnarray}
We have defined $\tau^2\equiv {\tilde p}^2\, T^2 =
(\theta_{\mu\nu}p^\nu)^2\, T^2$. However, in view of condition
(\ref{24}), this parameter is actually independent of the energy $p_0$.

As  $G_{\mu\nu}$ in Eq. (\ref{17}) is homogeneous of degree zero in $k$, 
$\Pi_{\mu\nu}^{AB}$ acquires to leading order an overall factor of $T^2$.

We now consider the various limits referred to in the
introduction. With the limit $\theta\, p\, T\gg 1$, 
we see that all terms proportional to $\delta_{A,0}$ vanish, 
so that non-planar graphs no longer contribute. This
leaves us with some straightforward integration that leads to
\be
\Pi^{\mu\;AB}_{\;\;\mu}  = N\,\delta^{A,B}\,{g^2\,T^2\over 3},
\label{29}\ee
\be
\Pi_{00}^{AB} = N\, \,\delta^{A,B}\,{g^2\,T^2\over 3}
\left(1-{\pop\over 2}\log{{\pop+1\over\pop-1}}\right)
\label{30}\ee
and
\be
{\tilde p^\mu \tilde p^\nu\over \tilde p^2} \Pi_{\mu\nu}^{AB}
=  -N\delta^{A,B}\,{g^2\,T^2\over 3}
{\pop\over 2}\left[{1\over 2}(\pop^2-1)
\log{{\pop+1\over\pop-1}}-\pop\right]
\label{31}\ee
Furthermore, in the limit $\theta\, p\, T\ll 1$, it is possible to extract
the contribution to the sums in Eqs. (\ref{26}) to (\ref{28}) that are
of leading order in $\tpt^2$. Using the standard
results for the Riemann zeta function, $\zeta(2) = \pi^2/6$ and
$\zeta(4) = \pi^4/90$, we find that
\be
\Pi^{\mu\;AB}_{\;\;\mu}  \approx  N\,\delta^{A,B}\,{g^2\,T^2\over 3} 
\left[1 -  \delta^{A,0} + {(\pi\,\tpt)^2\over 15}
\delta^{A,0}\right],
\label{32}\ee
\begin{eqnarray}
\Pi^{AB}_{00} & \approx & N\,\delta^{A,B}\,{g^2\,T^2\over 3}\left[
\left(1-{\pop\over 2}\log{{\pop+1\over\pop-1}}\right)(1-\delta^{A,0})
\right. \nnbb
& + & \left. {(\pi\tpt)^2\over 15}
\left(2-3\pop^2+{3\over 2}\pop(\pop^2-1)\log{{\pop+1\over\pop-1}}\right)
\delta^{A,0}\right]
\label{33}\end{eqnarray}
and
\begin{eqnarray}
{\tilde p^\mu \tilde p^\nu\over \tilde p^2} \Pi_{\mu\nu}^{AB}
& = & -N\,\delta^{A,B}\,\,{g^2\, T^2\over 3}
\left\{{\pop\over 2}\left[{1\over 2}(\pop^2-1)
\log{{\pop+1\over\pop-1}}-\pop\right]
\left(1-\delta^{A,0}\right)\right.\nonumber\\
& + & \left.{(\pi\,\tpt)^2\over 15}
\left[1-{15\over 4}\pop^2+{9\over 4}\pop^4
-{9\over 8}\pop\left(\pop^2-1\right)^2
\log{{\pop+1\over\pop-1}}\right]\delta^{A,0}\right\}.
\nnbb 
\label{34}\end{eqnarray}
Terms of order $\tpt^4$ and beyond can be similarly computed. We
note that all terms involving $\theta$ in Eqs. (\ref{32}) to
(\ref{34}) are proportional to $\delta^{A,0}$.

The integration over $\zeta$ in Eqs. (\ref{27}) and (\ref{28})
can be performed and in the static limit when $\pop,\,p_0\rightarrow
0$. We find in this case that
\be
\lim_{\pop\rightarrow 0} \Pi_{00}^{AB} =
N\,\delta^{A,B}\, {g^2\, T^2\over 3}\left[1-{6\over\pi^2} 
\sum_{n=1}^{\infty}{n^2\over (n^2+ \tpt^2)^2}\,\delta^{A,0}\right],
\label{35}\ee
\be
\lim_{\pop\rightarrow 0} 
{\tilde p^\mu \tilde p^\nu\over \tilde p^2} \Pi_{\mu\nu}^{AB} = 
- 2\,N\,\delta^{A,B}\, {g^2\, T^2\over \pi^2}\,\delta^{A,0}\,
\sum_{n=1}^{\infty}{\tpt^2\over (n^2+\tpt^2)^2}.
\label{36}\ee
(Of course, Eq. (\ref{26}) remains unchanged in the static limit.)
The sums appearing in Eqs. (\ref{26}), (\ref{35}) and (\ref{36}) are
standard \cite{gradshteyn}:
\be
\sum_{n=1}^\infty{1\over n^2 + a^2} = {\pi\over 2\, a}
{\rm coth}(\pi\, a) - {1\over 2\, a^2},
\label{37}\ee
\be
\sum_{n=1}^\infty{1\over (n^2 + a^2)^2} = {\pi\over 4\, a^3}
{\rm coth}(\pi\, a) + {\pi^2\over 4\, a^2}{\rm csch}^2(\pi \,a)
- {1\over 2\, a^4}.
\label{38}\ee
We consequently are led to 
\be
\Pi^{\mu\;AB}_{\;\;\mu}  = N\,\delta^{A,B}{g^2\,T^2\over 3} 
\left[1 -  3\left(
 {1\over \pi\,\tpt}\coth(\pi \tpt) -
{1\over (\pi\,\tpt)^2}\right)\delta^{A,0}\right],
\label{39}\ee
\be
\lim_{\pop\rightarrow 0} \Pi_{00}^{AB} = 
N\,\delta^{A,B}\, {g^2\, T^2\over 3}\left\{1-
{3\over 2}\left[{1\over \pi\tpt}\coth(\pi\tpt) -
{\rm csch}^2(\pi\tpt)\right]\delta^{A,0}\right\}
\label{40}\ee
and
\begin{eqnarray}
\lim_{\pop\rightarrow 0} 
{\tilde p^\mu \tilde p^\nu\over \tilde p^2} \Pi_{\mu\nu}^{AB} & = &
-N\,\delta^{A,B}\, {g^2\, T^2}
{{\rm csch}^2(\pi\tpt)\over 2\,(\pi\,\tpt)^2} 
\left[1 + (\pi\tpt)^2 - {\rm cosh}(2\pi\tpt) +
{\pi\tpt\over 2}{\rm sinh}(2\pi\tpt)\right]\,\delta^{A,0}.
\label{41}\end{eqnarray}

From the previous results we can now obtain the static limit of 
Eqs. (\ref{20}), (\ref{21}) and (\ref{22}) (we show in the appendix 
that $\Pi_4^{AB}=0$). From Eq. (\ref{20}), we have
\be
\lim_{\pop\rightarrow 0} \Pi_1^{AB} = 
\lim_{\pop\rightarrow 0} \Pi^{\mu\;\;AB}_{\;\;\mu} -
\lim_{\pop\rightarrow 0} \Pi^{AB}_{00} - 
\lim_{\pop\rightarrow 0} 
{\tilde p^\mu \tilde p^\nu\over \tilde p^2} \Pi_{\mu\nu}^{AB}.
\label{42}\ee
Using Eqs. (\ref{26}), (\ref{35}) and (\ref{36}), we
obtain 
\be
\lim_{\pop\rightarrow 0} \Pi_1^{AB} = 0.
\label{43}\ee
The static limit of $\Pi_2$ in Eq. (\ref{21}) behaves as
$\pop^2$. Therefore, $\Pi_2$ will contribute only to
$\Pi_{00}^{AB}$ in Eq. (\ref{19}) but not to either
$\Pi_{ij}^{AB}$ or $\Pi_{0i}^{AB}$ ($i,j=1,2,3$) [of course, we do not need
$\Pi_2$ in order to obtain $\Pi_{00}^{AB}$ which has already been
obtained explicitly in Eq. (\ref{40})].
Using Eq. (\ref{22}), we obtain for the
static limit of $\Pi_3$
\begin{eqnarray}
\lim_{\pop\rightarrow 0} \Pi_3^{AB} & = &
-{1\over |\vec{\tilde p}|^2 |{\vec p}|^2}
\lim_{\pop\rightarrow 0}
\left[\Pi^{\mu\;\;AB}_{\;\;\mu}
-\Pi_{00}^{AB}
-2{\tilde p^\mu \tilde p^\nu\over \tilde p^2} \Pi_{\mu\nu}^{AB}\right]
\nnbb & = & - 3 \lim_{\pop\rightarrow 0}
{\tilde p^\mu \tilde p^\nu\over \tilde p^2} \Pi_{\mu\nu}^{AB}
\label{44}\end{eqnarray}
Finally, using  Eq. (\ref{41}) in Eq. (\ref{44}) as well as simple 
functional relations involving the hyperbolic functions, we obtain
\be
\lim_{\pop\rightarrow 0} \Pi_3^{AB} = N\,\delta^{A,B}\,\delta^{A,0}
{3\,g^2\,T^2\over 2}
\left[
{1\over {\rm sinh}^2(\pi\,\tpt)} - {2\over (\pi\,\tpt)^2}
+  {{\rm cosh}(\pi\,\tpt)\over \pi\,\tpt {\rm sinh}(\pi\,\tpt)} 
\right].
\label{45}\ee

Inserting Eq. (\ref{45}) into Eq. (\ref{19}) one can easily obtain 
$\Pi_{i\,j}^{AB}$.
It is interesting to note that in commutative Yang-Mills theory
$\Pi_{00}^{AB}$ is non-zero but $\Pi_{ij}^{AB}$ vanishes, while in the
noncommutative case, we see from Eq. (\ref{45}) that 
$\Pi_{ij}^{AB}\neq 0$ when then colour $A$ is in the ${\rm U}(1)$
sector ($A=0$). This is consistent with the additional magnetic
interactions appearing in the initial Lagrangian. However, since 
$\Pi^{AB}_{ii}(p_0=0,\vec p\rightarrow 0)=0$, the magnetic mass
vanishes also in the noncommutative theory.

It also proves possible to examine $\Pi_{\mu\nu}^{AB}$ in the long
wave length limit $\sigma\rightarrow\infty$ and $\tau\rightarrow 0$. 
In this case, we see from Eqs. (\ref{26}), (\ref{27}) and (\ref{28}) that
\begin{eqnarray}
\Pi_{00}^{AB} &\longrightarrow & 0, \nnbb
\Pi_{\mu}^{AB\,\mu} &\longrightarrow & N\delta^{A,B}
{g^2\,T^2\over 3}\left(1-\delta^{A,0}\right),\nnbb
{\tilde p^\mu \tilde p^\nu\over {\tilde p}^2}
\Pi_{\mu\nu}^{AB} &\longrightarrow & N\delta^{A,B}
{g^2\,T^2\over 9}\left(1-\delta^{A,0}\right).
\end{eqnarray}

\section{The three-point function}\label{sec3}

\begin{figure}[h!]
\includegraphics*{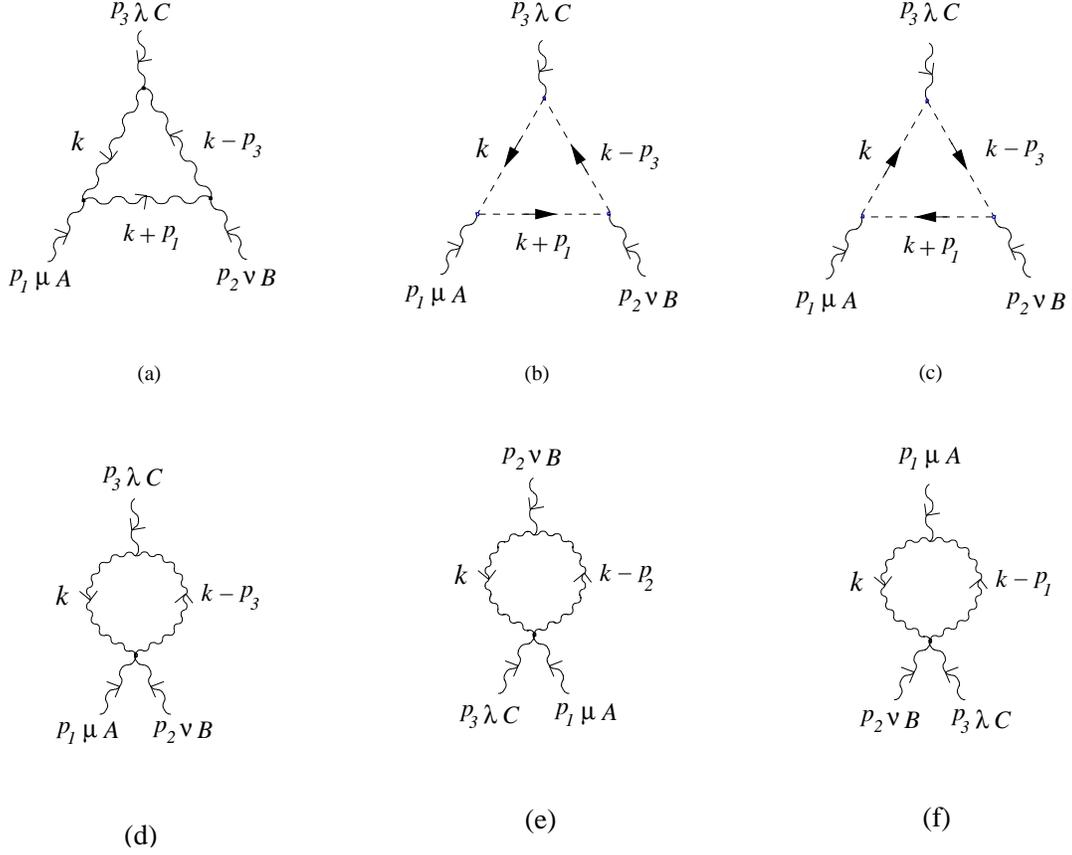}
\caption{One-loop diagrams which contribute to the three-point function.}
\label{fig3}
  \end{figure}

\begin{figure}[h!]
\includegraphics*{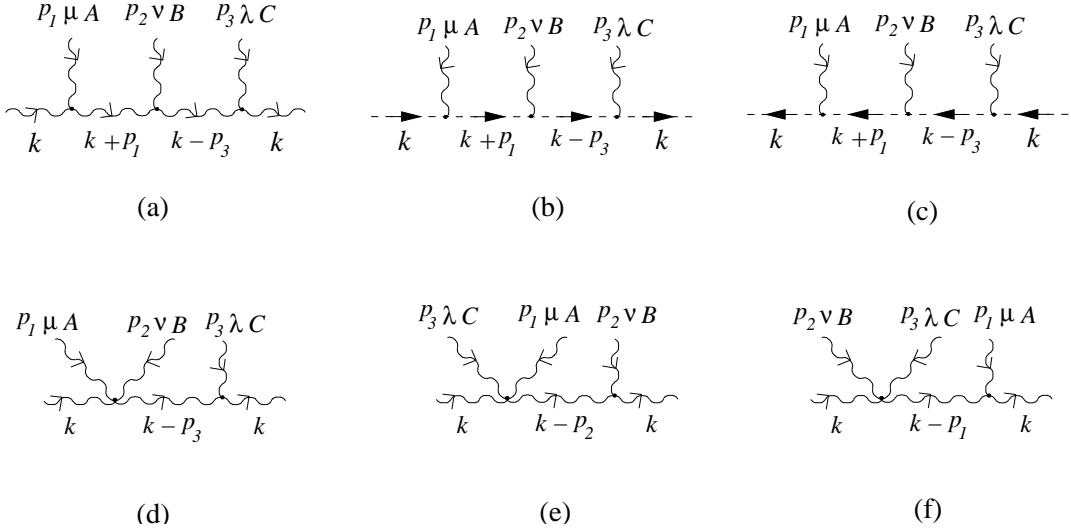}
\caption{Forward scattering amplitudes corresponding to the
diagrams in Figure \ref{fig3}. The direction of the ghost momentum is the same
as the corresponding internal gluon line. 
Permutations of the vertices are understood.}
\label{fig4}\end{figure}

The Feynman graphs we consider are those of Fig. \ref{fig3}. These are
associated with the 
amplitudes of Fig. \ref{fig4}.
The calculation of the diagrams in figure \ref{fig4} is
straightforward but very tedious.
After some algebra, the amplitude in figure \ref{fig4} (a) 
can be written as (we have employed the Maple version of the
symbolic computer package HIP \cite{hsieh:1992ti})
\be
{\cal A}_{\mu\nu\lambda}^{gl\;ABC} =
\left[\sin\left({p_1\times p_2\over 2}\right)C_{\sin}^{gl\;ABC} +
      \cos\left({p_1\times p_2\over 2}\right)C_{\cos}^{gl\;ABC}
\right]L^{gl}_{\mu\nu\lambda}
\label{46}\ee
Similarly the sum of the diagrams in figures \ref{fig4} (b) and 
\ref{fig4} (c) give
\be
{\cal A}_{\mu\nu\lambda}^{gh\;ABC} =
\left[\sin\left({p_1\times p_2\over 2}\right)C_{\sin}^{gh\;ABC} +
      \cos\left({p_1\times p_2\over 2}\right)C_{\cos}^{gh\;ABC}
\right]L^{gh}_{\mu\nu\lambda}
\label{47}\ee

The factors $C_{\sin}^{gl\;ABC}$, $C_{\cos}^{gl\;ABC}$, $C_{\sin}^{gh\;ABC}$ and 
$C_{\cos}^{gh\;ABC}$ are trigonometric functions of the internal momentum 
$k$ and involve the colour factors. At any specific order in the hard 
thermal loop expansion, the Lorentz factors $L^{gl}_{\mu\nu\lambda}$ and  
$L^{gh}_{\mu\nu\lambda}$ will be odd or even in $k$. Terms which are
odd in $k$ will be multiplied by the following antisymmetric factors
\begin{eqnarray}
A_{\sin}^{gl\;ABC} = {1\over 2}\left(C_{\sin}^{gl\;ABC}(k)-C_{\sin}^{gl\;ABC}(-k)\right) 
\nnbb
A_{\sin}^{gh\;ABC} = {1\over 2}\left(C_{\sin}^{gh\;ABC}(k)-C_{\sin}^{gh\;ABC}(-k)\right)
\nnbb
A_{\cos}^{gl\;ABC} = {1\over 2}\left(C_{\cos}^{gl\;ABC}(k)-C_{\cos}^{gl\;ABC}(-k)\right) 
\nnbb
A_{\cos}^{gh\;ABC} = {1\over 2}\left(C_{\cos}^{gh\;ABC}(k)-C_{\cos}^{gh\;ABC}(-k)\right)
\label{48}\end{eqnarray}
Terms which are even in $k$ will be multiplied by the 
following symmetric factors
\begin{eqnarray}
S_{\sin}^{gl\;ABC} = {1\over 2}\left(C_{\sin}^{gl\;ABC}(k)+C_{\sin}^{gl\;ABC}(-k)\right) 
\nnbb
S_{\sin}^{gh\;ABC} = {1\over 2}\left(C_{\sin}^{gh\;ABC}(k)+C_{\sin}^{gh\;ABC}(-k)\right)
\nnbb
S_{\cos}^{gl\;ABC} = {1\over 2}\left(C_{\cos}^{gl\;ABC}(k)+C_{\cos}^{gl\;ABC}(-k)\right) 
\nnbb
S_{\cos}^{gh\;ABC} = {1\over 2}\left(C_{\cos}^{gh\;ABC}(k)+C_{\cos}^{gh\;ABC}(-k)\right)
\label{49}\end{eqnarray}

A straightforward calculation gives
\begin{eqnarray}
A_{\sin}^{gl\;ABC} & = &  -\tr\left[F^A\,D^B\,D^C\,c_{{1\over
      2}p_1}\,c_{{1\over 2}p_2}\,\,s_{{1\over 2}p_3}\,
                                   -F^A\,F^B\,F^C\,c_{{1\over
                                       2}p_1}\,s_{{1\over
                                       2}p_2}\,\,c_{{1\over 2}p_3}\,
\right.
\nnbb
& & 
\;\;\;\;\;\left.
                                   +F^C\,D^A\,D^B\,s_{{1\over
                                       2}p_1}\,\,c_{{1\over
                                       2}p_2}\,\,c_{{1\over 2}p_3}\,
                                   -F^B\,D^C\,D^A\,s_{{1\over 2}p_1}\,\,s_{{1\over 2}p_2}\,\,s_{{1\over 2}p_3}\,\right],
\label{50}\end{eqnarray}
where we are using the abbreviations defined in Eqs. (\ref{7}) and (\ref{8}).
Using the relations \cite{Armoni:2000xrBonora:2000ga}
\be
\tr F^A\,F^B\,F^C = -{N\over 2} f^{ABC} = - \tr F^A\,D^B\,D^C,
\label{51}\ee
as well as the cyclic property of the trace, we can write
\begin{eqnarray}
A_{\sin}^{gl\;ABC} & = &  -{N\over 2}f^{ABC}\left[\,c_{{1\over 2}p_1}\,c_{{1\over 2}p_2}\,\,s_{{1\over 2}p_3}\,
                                             +\,c_{{1\over 2}p_1}\,s_{{1\over 2}p_2}\,\,c_{{1\over 2}p_3}\,\right.
\nnbb
& & 
\;\;\;\;\;\;\;\;\;\;\;\;\;\;\left.
                                   +\,s_{{1\over 2}p_1}\,\,c_{{1\over 2}p_2}\,\,c_{{1\over 2}p_3}\,\
                                   -\,s_{{1\over 2}p_1}\,\,s_{{1\over 2}p_2}\,\,s_{{1\over 2}p_3}\,\right]
\nnbb
&= &  -{N\over 2}f^{ABC} \sin\left({p_1+p_2+p_3\over 2}\times k\right)
                                             = 0,
\label{52}\end{eqnarray}
where we have used the momentum conservation $p_1+p_2+p_3=0$.
A similar calculation for the ghost part also gives
\be
A_{\sin}^{gh\;ABC} = 0.
\label{53}\ee
The Eqs. (\ref{52}) and (\ref{53}) imply that there is no contribution 
which is odd in the temperature $T$ {\it and} proportional to 
$\sin[(p_1\times p_2)/2]$. 

Let us now consider the coefficients of  $\cos[(p_1\times p_2)/2]$
which are antisymmetric functions of $k$, namely  $A_{\cos}^{gl\;ABC}$ 
and $A_{\cos}^{gh\;ABC}$. The colour traces which are involved now are
\cite{Armoni:2000xrBonora:2000ga}
\begin{eqnarray}
\tr D^A D^B D^C & = & {N\over 2}\eta_{ABC}\,d^{ABC};\;\;\;
\eta_{ABC}\equiv d_A d_B d_C - 4\delta_{A+B+C,0}
\nnbb
\tr F^A F^B D^C & = & -{N\over 2}c_A c_B d_C\,d^{ABC};\;\;
c_A\equiv 1 - \delta_{A,0};\;\; d_A\equiv 1 + \delta_{A,0}
\label{54}\end{eqnarray}
A straightforward calculation gives
\begin{eqnarray}
A_{\cos}^{gl\;ABC} & = &  -\tr\left[F^A\,F^B\,D^C\,c_{{1\over 2}p_1}\,c_{{1\over 2}p_2}\,\,s_{{1\over 2}p_3}\,\right.
\left.
                              +D^A\,F^B\,F^C\,s_{{1\over 2}p_1}\,\,c_{{1\over 2}p_2}\,\,c_{{1\over 2}p_3}\,\right.
\nnbb
& & 
\;\;\;\;\;\left.
                              +F^C\,F^A\,D^B\,c_{{1\over
                              2}p_1}\,s_{{1\over 2}p_2}\,\,c_{{1\over
                              2}p_3}\,
                              +D^A\,D^B\,D^C\,s_{{1\over 2}p_1}\,\,s_{{1\over 2}p_2}\,\,s_{{1\over 2}p_3}\,\right]
\label{55}\end{eqnarray}
In contrast to the antisymmetric coefficient of $\sin[(p_1\times p_2)/2]$, 
$A_{\cos}^{gl\;ABC}$ does not vanish by itself. Using trace cyclicity,
one can write 
\begin{eqnarray}
A_{\cos}^{gl\;ABC} & = &  
\tr\left[-F^A\,F^B\,D^C\,c_{{1\over 2}p_1}\,c_{{1\over 2}p_2}\,\,s_{{1\over 2}p_3}\, + {1\over 3}
D^A\,D^B\,D^C\,s_{{1\over 2}p_1}\,\,s_{{1\over 2}p_2}\,\,s_{{1\over 2}p_3}\, \right] \nnbb 
& + &  {\rm cyclic\;\;\; permut.}
\label{56}\end{eqnarray}
Let us consider the
specific case of the {\it superleading} contribution, which would be
proportional to $T^3$. The Lorentz factor for this piece is
the same for the gluon and for the ghost diagrams, being proportional to  
[see the diagrams in figure \ref{fig4} (a), (b) and (c)],
\be
{k_\mu k_\nu k_\lambda\over k\cdot p_1 k\cdot p_3}
\label{57}\ee
Including the colour and the $\cos\left({p_1\times p_2\over 2}\right)$
factors, this leads to a contribution
\be
\cos\left({p_1\times p_2\over 2}\right)A_{\cos}^{gl\;ABC}(p_1,p_2,p_3)
k_\mu k_\nu k_\lambda\,{1\over k\cdot p_1 k\cdot p_3}
\label{58}\ee
The only part of Eq. (\ref{58}) which will change when we add together
all cyclic permutations is the denominator. Using momentum
conservation, we have
\be
  {1\over k\cdot p_1 \;k\cdot p_3} + {1\over k\cdot p_1 \;k\cdot p_2}
+ {1\over k\cdot p_2 \;k\cdot p_3} = 0,
\label{59}\ee
and thus we can easily see that the superleading contribution 
(viz. those that are proportional to $T^3$) will vanish.
The same property is also true for the ghost diagram. This
cancellation of the superleading contribution is similar to what
happens in QCD \cite{frenkel:1991ts}. However, in the case of noncommutative
theories one has to employ the cyclicity property of the
colour/trigonometric factor, rather than simply relying in the
cancellation which occurs when we add the contributions with
$k\rightarrow -k$, as is the case in QCD. (In QCD, there is no
$k$-dependent trigonometric factor which is odd in $k$.)

In a similar way, it is straightforward to show that
\begin{eqnarray}
S_{\cos}^{gl\;ABC} & = &  \tr\left[
F^A\,F^B\,F^C\,c_{{1\over 2}p_1}\,c_{{1\over 2}p_2}\,\,c_{{1\over 2}p_3}
+
F^A\,D^B\,D^C\,c_{{1\over 2}p_1}\,\,s_{{1\over 2}p_2}\,\,s_{{1\over 2}p_3}\,\right.
\nnbb
& & 
\;\;\;\;\;\left.
+
F^B\,D^C\,D^A\,s_{{1\over 2}p_1}\,c_{{1\over 2}p_2}\,\,s_{{1\over 2}p_3}
+
F^C\,D^A\,D^B\,s_{{1\over 2}p_1}\,\,s_{{1\over 2}p_2}\,\,c_{{1\over 2}p_3}\,\right]
\nnbb
& = & -{N\over 2}f^{ABC} \cos\left[{1\over 2}(p_1+p_2+p_3)\times k\right] = 
-{N\over 2}f^{ABC},
\label{60}\end{eqnarray}
\begin{eqnarray}
S_{\cos}^{gh\;ABC} & = &  \tr\left[
F^A\,F^B\,F^C\left(c_{{1\over 2}p_1}^2\,c_{{1\over 2}p_3}^2
-c_{{1\over 2}p_1}\, c_{{1\over 2}p_3}
 s_{{1\over 2}p_1}\, s_{{1\over 2}p_3}\right)
-F^A\,D^B\,D^C\left(c_{{1\over 2}p_1}^2\,\,s_{{1\over 2}p_3}^2
+c_{{1\over 2}p_1}\, c_{{1\over 2}p_3}
 s_{{1\over 2}p_1}\, s_{{1\over 2}p_3}\right)\right.
\nnbb
& & \left.
-F^B\,D^C\,D^A\left(s_{{1\over 2}p_1}^2\,s_{{1\over 2}p_3}^2
-c_{{1\over 2}p_1}\, c_{{1\over 2}p_3}
 s_{{1\over 2}p_1}\, s_{{1\over 2}p_3}\right)
-F^C\,D^A\,D^B\left(s_{{1\over 2}p_1}^2\,\,c_{{1\over 2}p_3}^2
+c_{{1\over 2}p_1}\, c_{{1\over 2}p_3}
 s_{{1\over 2}p_1}\, s_{{1\over 2}p_3}\right)
\right]
\nnbb
& = & 
-{N\over 2}f^{ABC},
\label{61}\end{eqnarray}
\begin{eqnarray}
S_{\sin}^{gl\;ABC} & = &  \tr\left[-
F^A\,F^B\,D^C\,c_{{1\over 2}p_1}\,s_{{1\over 2}p_2}\,\,s_{{1\over 2}p_3}
-
F^B\,F^C\,D^A\,s_{{1\over 2}p_1}\,\,s_{{1\over 2}p_2}\,\,c_{{1\over 2}p_3}\,\right.
\nnbb
& & 
\;\;\;\;\;\left.
+
F^C\,F^A\,D^B\,c_{{1\over 2}p_1}\,c_{{1\over 2}p_2}\,\,c_{{1\over 2}p_3}
+
D^A\,D^B\,D^C\,s_{{1\over 2}p_1}\,\,c_{{1\over 2}p_2}\,\,s_{{1\over 2}p_3}\,\right]
\nnbb
& = & -{N\over 2}d^{ABC} \left[
1+
c_{p_1}\left(\delta^{A,0}\delta^{B,0}\delta^{C,0}-
\delta^{B,0}\delta^{C,0} - \delta^{A,0} \right)\right.
\nnbb &  & 
\;\;\;\;\;\;\;\;\;\;\;\;\;\;\;\;\;\;\; - \; 
c_{p_2}\left(\delta^{A,0}\delta^{B,0}\delta^{C,0}-
\delta^{A,0}\delta^{C,0} - \delta^{B,0} \right)
\nnbb &  & 
\;\;\;\;\;\;\;\;\;\;\;\;\;\;\;\;\;\;\; \left. - \;
c_{p_3}\left(\delta^{A,0}\delta^{B,0}\delta^{C,0}-
\delta^{A,0}\delta^{B,0} - \delta^{C,0} \right)
\right],
\label{62}\end{eqnarray}
and
\begin{eqnarray}
S_{\sin}^{gh\;ABC} & = &  {1\over 4}
\tr\left[F^A\,F^B\,D^C + F^A\,D^B\,F^C + D^A\,F^B\,F^C - D^A\,D^B\,D^C 
\right.\nnbb
\nnbb
& & 
\;\;\;\;\;
+ c_{p_1}
\left(F^A\,F^B\,D^C + F^A\,D^B\,F^C - D^A\,F^B\,F^C + D^A\,D^B\,D^C \right)
\nnbb
& & 
\;\;\;\;\;
+ c_{p_2}
\left(-F^A\,F^B\,D^C + F^A\,D^B\,F^C - D^A\,F^B\,F^C - D^A\,D^B\,D^C \right)
\nnbb
& & 
\;\;\;\;\; \left.
+ c_{p_3}
\left(-F^A\,F^B\,D^C + F^A\,D^B\,F^C + D^A\,F^B\,F^C + D^A\,D^B\,D^C \right)
\right] \nnbb
& = & -{N\over 2}d^{ABC} \left[
1+
c_{p_1}\left(\delta^{A,0}\delta^{B,0}\delta^{C,0}-
\delta^{B,0}\delta^{C,0} - \delta^{A,0} \right)\right.
\nnbb &  & 
\;\;\;\;\;\;\;\;\;\;\;\;\;\;\;\;\;\;\; - \; 
c_{p_2}\left(\delta^{A,0}\delta^{B,0}\delta^{C,0}-
\delta^{A,0}\delta^{C,0} - \delta^{B,0} \right)
\nnbb &  & 
\;\;\;\;\;\;\;\;\;\;\;\;\;\;\;\;\;\;\; \left. + \;
c_{p_3}\left(\delta^{A,0}\delta^{B,0}\delta^{C,0}-
\delta^{A,0}\delta^{B,0} - \delta^{C,0} \right)
\right].
\label{63}\end{eqnarray}

By Eqs. (\ref{60}) to (\ref{63}) it is evident that the full 
amplitude associated with Figs. \ref{fig4} (a), (b) and (c) is
\begin{eqnarray}
\left.{\cal A}_{\mu\nu\lambda}^{ABC}\right|_{(a),(b),(c)}
& = &
-{N\over 2}\left[f^{ABC} \cos\left({p_1\times p_2\over 2}\right) +
d^{ABC}\left(1+O^{ABC}\right) 
\sin\left({p_1\times p_2\over 2}\right)\right]
\nnbb & \times & \left[L^{gl}_{\mu\nu\lambda} + L^{gh}_{\mu\nu\lambda}\right],
\label{64}\end{eqnarray}
where 
\begin{eqnarray}
O^{ABC} & = & \cos(p_1\times k)\left(\delta^{A,0}\delta^{B,0}\delta^{C,0}-
\delta^{B,0}\delta^{C,0} - \delta^{A,0} \right)\nnbb
& - & \cos(p_2\times k)\left(\delta^{A,0}\delta^{B,0}\delta^{C,0}-
\delta^{A,0}\delta^{C,0} - \delta^{B,0} \right)\nnbb
& + & \cos(p_3\times k)\left(\delta^{A,0}\delta^{B,0}\delta^{C,0}-
\delta^{A,0}\delta^{B,0} - \delta^{C,0} \right)
\label{65}\end{eqnarray}
is an oscillatory part which gives a subleading contribution for 
$\theta\,p\,T\gg 1$. Furthermore, explicit computation gives
\begin{eqnarray}
L^{gl}_{\mu\nu\lambda} + L^{gh}_{\mu\nu\lambda} & = &
{i\over k\cdot p_3}\left[
{{4\,{{p_1}}^{2}k_{{\mu}}k_{{\nu}}k_{{\lambda}}}\over
{{{(k\cdot p_1)}}^{2}}}+{{4\,k_{{\nu}}k_{{\mu}}
{p_3}_{{\lambda}}}\over{{k\cdot p_1}}}+
{{4\,{p_3}_{{\nu}}k_{{\mu}}k_{{\lambda}}}\over{{k\cdot p_1}}}-
k_{{\mu}}\eta_{{\lambda,\nu}}-k_{{\nu}}\eta_{{\lambda,\mu}}\right] \nnbb
& - & \left([(\mu,p_1),(\lambda,p_3)]\right) \longleftrightarrow
\left([(\lambda,p_3),(\mu,p_1)]\right). 
\label{66}\end{eqnarray}
(As in the case of the two-point function, in the leading thermal contributions,
all dependence of the three-point function on the gauge parameter 
$\xi$ cancels completely.)
The full contribution to the three-point function is obtained adding
to Eq. (\ref{64}) two cyclic permutations of 
$(\mu,p_1),\;(\nu,p_2)\;(\lambda,p_3)$. In the limit 
$\theta\,p\,T\gg 1$, when $O^{ABC}$ can be neglected, the
colour/trigonometric factor in Eq. (\ref{64}) does not change under
cyclic permutations. Therefore, we can write
\begin{eqnarray}
\lim_{(\theta\,p\,T) \rightarrow\infty}\left[
\left.{\cal A}_{\mu\nu\lambda}^{ABC}\right|_{(a),(b),(c)}
+ {\rm cyclic\;\; perm.\;\; of\;}(\mu,p_1),\;(\nu,p_2)\;(\lambda,p_3) \right]
& = & -{N\over 2}C^{ABC}(p_1,p_2)
\left[L^{gl}_{\mu\nu\lambda} + L^{gh}_{\mu\nu\lambda}\right]\nnbb
&+& {\rm cyclic\;\; perm.\;\; of\;}(\mu,p_1),\;(\nu,p_2)\;(\lambda,p_3),
\label{66a}\end{eqnarray}
where $C^{ABC}(p_1,p_2)$ is given by Eq. (\ref{apa3}).
Since the factor $C^{ABC}(p_1,p_2)$ does not change under cyclic permutations,
the Lorentz factor
$L^{gl}_{\mu\nu\lambda}+L^{gh}_{\mu\nu\lambda}$,
plus its cyclic permutations simplifies to an expression without
terms involving the metric tensor $\eta$, so that the full expression
from the diagrams in Figs. \ref{fig4} (a), (b) and (c) can be written as
\begin{eqnarray}
\lim_{(\theta\,p\,T) \rightarrow\infty}
\left.{\cal A}_{\mu\nu\lambda}^{ABC}\right|_{(a),(b),(c)}
&=& -{i\, N\,C^{ABC}(p_1,p_2)\over k\cdot p_3}
\left[{k_\mu\,k_\nu\,k_\lambda\,p_3^2\over(k\cdot p_1)^2}
+ 2 {k_\mu\,k_\lambda\,{p_3}_\nu\over k\cdot p_1}
+ (\mu\leftrightarrow \nu)\right] \nnbb
&+&\left[(A,\mu,p_1),(C,\lambda,p_3)\right]\longleftrightarrow
\left[(C,\lambda,p_3),(A,\mu,p_1)\right].
\label{66b}\end{eqnarray}

The remaining contribution, associated with the 
amplitudes
in Figs. \ref{fig4} (d), (e) and (f),  is purely oscillatory and hence 
will not contribute when $\theta\, p\, T\gg 1$.
Including the two identical contributions which arise as a result of reversing the
momentum flow of $k$ in Fig. \ref{fig4} (d), (e), and (f) and also by 
interchanging the two vertices appearing there, we obtain
\begin{eqnarray}
\left.{\cal A}_{\mu\nu\lambda}^{ABC}\right|_{(d),(e),(f)}
&=&
-{2\,i\,N}\, d^{ABC}\sin\left({p_1\times p_2\over 2}\right)
\cos(p_1\times k)\left(\delta^{A,0}\delta^{B,0}\delta^{C,0}-
\delta^{B,0}\delta^{C,0} - \delta^{A,0} \right)\nnbb&\times&
\displaystyle{
{k_\mu\,\eta_{\nu\lambda} +k_\nu\,\eta_{\mu\lambda} +
4\,k_\lambda\,\eta_{\mu\nu}\over k\cdot p_3}}\nnbb
& + &  \left([(\mu,p_1),(\nu,p_2)]\right) \longleftrightarrow
       \left([(\nu,p_2),(\mu,p_1)]\right) \nnbb
& + & {\rm cyclic\;\; perm.\;\; of\;}
(\mu,p_1),\;(\nu,p_2)\;(\lambda,p_3).
\label{67}\end{eqnarray}

After using (\ref{64}) and (\ref{66}) in conjunction with
\be
\Gamma_{\mu\nu\lambda}^{ABC}(p_1,p_2,p_3) = 
-{1\over (2\pi)^3}\int{d^3\vec k\over 2\,|\vec k|} N(|\vec k|)
\left.{\cal A}_{\mu\nu\lambda}^{ABC}\right|_{k0=|\vec k|};\;\;\;\; 
A,B,C = 0,1,2,\cdots,N^2-1
\label{68}\ee
to compute the three-point function in the hard thermal limit, one is
confronted with very complicated angular integrals.
However, it is apparent that because ${\cal A}_{\mu\nu\lambda}^{ABC}(k)$ in
Eq. (\ref{68}) is homogeneous of zero degree in $k$, the three-point function
is quadratic in $T$, as in commutative Yang-Mills theory. Furthermore,
the simple Ward identity
\be
p_3^\lambda\,\Gamma_{\mu\nu\lambda}^{ABC}(p_1,p_2,p_3)=i\,
C^{ABX}(p_1,p_2)\,
\left(\Pi_{\mu\nu}^{XC}(p_1) - \Pi_{\nu\mu}^{XC}(p_2)\right)
\label{69}\ee
can be seen to be satisfied in the hard thermal limit when
$\theta\,p\,T\gg 1$, without having to perform the integration over
$\vec k$. This can be verified directly, at the integrand level,
using the explicit forms of the two- and three-point amplitudes given
by Eqs. (\ref{16}) (with $\theta\,p\,T\gg 1$) and (\ref{68}).
Actually, this Ward identity should be satisfied for all values of
$\theta\, p \, T$. This is because in the hard thermal limit
amplitudes with external ghost lines do not have a $T^2$ behaviour and
hence BRS identities reduce to simple Ward identities such as those in 
(\ref{69}).
The above Ward identity, together with the results given in 
Eqs. (\ref{40}) and (\ref{45}), implies that the leading $T^2$
contributions to the static three point amplitude are
non-vanishing. This behaviour contrasts with the one in the commutative
theory, where the gluon self-energy is the only static amplitude
with a hard thermal loop.

\section{Discussion}\label{sec4}
An essential ingredient of the resumation program is the computation
of the effective action for the hard thermal loops. In this work, we
have addressed the problem of obtaining the two- and three-point
functions in noncommutative \un Yang-Mills theory at high
temperature. These calculation are much more difficult than the
corresponding ones in commutative \sun theory, due to the presence of
the tensor $\theta_{\mu\nu}$, which appears in the interaction
vertices. It is interesting to remark at this point that, as
$\theta_{\mu\nu}\rightarrow 0$, the ${\rm U}(1)$ sector decouples,
so that the usual results of the \sun gauge theory are recovered. This
can be understood by noting that the
temperature does provide a natural ultraviolet cut-off for the thermal 
part of the amplitudes (in contrast, such limit is singular at $T=0$,
due to the phenomenon of the UV/IR mixing).
This fact enables one to take, for the leading 
thermal contributions in the noncommutative theory, the limit
$\theta_{\mu\nu}\rightarrow 0$ in a smooth way.

The approach which relates the hard thermal loops to the angular
integrals of forward scattering amplitudes of on-shell thermal
particles, allows one to infer much useful information about their
high-temperature behaviour.
From an examination of these amplitudes,
where the leading terms are all of order $T^2$, one learns the
following properties of the angular integrands:

\begin{itemize}
\item[(a)] The non-localities involve, in configuration space, products of
$\left(k\cdot\partial\right)^{-1}$.

\item[(b)] Apart from the trigonometric factors involving the noncommutative
parameter, the integrands are homogeneous functions of $k$ of zero
degree and Lorentz covariant.

\item[(c)] They are gauge invariant and satisfy simple Ward identities
analogous to those of the tree amplitudes.

\end{itemize}

Using similar arguments to
the ones employed in reference \cite{brandt:1993mj}, we expect that
these properties, together with the results for the lowest order
amplitudes, may be sufficient to determine the effective action
for hard thermal loops. This issue of the noncommutative Yang-Mills theory
is currently being considered.

\begin{acknowledgments}
We would  like to thank Professors A. Das and J. C. Taylor for
helpful discussions.  D.G.C. McKeon would like to thank the 
Universidade de S\~ao Paulo for the hospitality and R. and D. 
MacKenzie for encouragement. This work was supported by CNPq 
and FAPESP, Brazil. 
\end{acknowledgments}

\appendix

\section{Feynman rules}\label{appA}

The propagators of the gauge and ghost particles are respectively
given by
\begin{eqnarray}
\includegraphics*{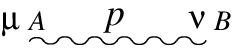}
& : \;\;\;&\;\;\; -\displaystyle{
{i\,\,\delta^{A,B}\,\over (p^{2}+i\epsilon)} \left(\eta_{\mu\nu} -
 (1-\xi){p_{\mu}p_{\nu}\over p^{2}}\right)} \nnbb
\includegraphics*{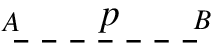}
 & : \;\;\;&\;\;\; 
\displaystyle{{i\,\,\delta^{A,B}\,\over p^{2} + i\epsilon}}
\label{apa1}\end{eqnarray}
The vertices are
\begin{eqnarray}
\begin{array}{c}\includegraphics*{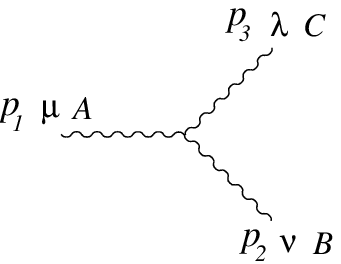}\end{array}
 & :\;\;\;  &\;\;\;  -g\,C^{ABC}(p_1,p_2)
\left[(p_{1}-p_{2})^{\lambda}\eta^{\mu\nu} +
      (p_{2}-p_{3})^{\mu}\eta^{\nu\lambda} +
      (p_{3}-p_{1})^{\nu}\eta^{\lambda\mu}\right]\nnbb  & & \nnbb
\begin{array}{c}\includegraphics*{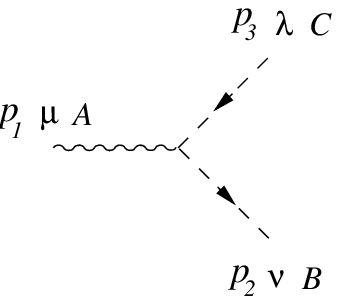}\end{array}
 & :\;\;\;  &\;\;\;  g\,C^{ABC}(p_2,p_3)\, p_{2}^{\mu}
 \nnbb & & \nnbb  
\begin{array}{c}\includegraphics*{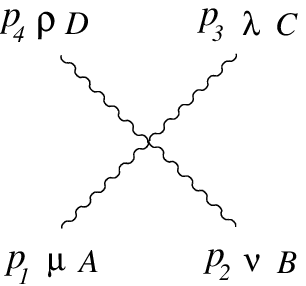}\end{array}
 & :\;\;\;  &\;\;\;  -i\, g^{2}\left[
C^{ABX}(p_1,p_2)\;C^{XCD}(p_3,p_4)
(\eta^{\mu\lambda}\eta^{\nu\rho} - \eta^{\mu\rho}\eta^{\nu\lambda}) 
\right.\nonumber\\
 &  & \qquad\quad + 
C^{ACX}(p_1,p_3)\;C^{XBD}(p_4,p_2)
(\eta^{\mu\rho}\eta^{\lambda\nu} - \eta^{\mu\nu}\eta^{\lambda\rho}) 
\nonumber\\ & & \nonumber\\ & & \nonumber\\
 &  & \qquad\quad\left. + 
C^{ADX}(p_1,p_4)\;C^{XBC}(p_2,p_3)
(\eta^{\mu\nu}\eta^{\lambda\rho} - \eta^{\mu\lambda}\eta^{\nu\rho}) 
\right],\nnbb & & \label{vertex}
\end{eqnarray}
where
\be
C^{ABC}(p_1,p_2) = f^{ABC}\,\cos({p_1\times p_2\over 2}) +
                   d^{ABC}\,\sin({p_1\times p_2\over 2});\;\;\;\;
p_1\times p_2\equiv p_1^\mu \theta_{\mu\nu} p_2^\nu
\label{apa3}\ee
and {\it all momenta} are inward. Dirac delta functions for the
conservation of momenta are understood.

\section{Integrals}\label{appB}
In this appendix, the integrals appearing in Eqs. (\ref{20}) to
(\ref{23}) are evaluated, leaving us in general with answers in terms
of infinite sums.

First of all, from Eqs. (\ref{4}), (\ref{16}) and (\ref{17}) we obtain
\be
p_0 \tilde p^\mu \Pi_{\mu 0}^{AB} = {2g^2N\,\delta^{A,B}\,\over(2\pi)^3}
\int{d^3\vec k\over |\vec k|} N(|\vec k|)
\left[1-\delta^{A,0}\cos(k\cdot\tilde p)\right] K\cdot\tilde p
\left[{p^2\over (K\cdot p)^2} - {p_0\over K\cdot p}\right]
\label{b1};\;\;\; K\equiv\left(1,{\vec k\over |\vec k|}\right)\ee
One can perform the previous integral using a coordinate system such
that two of the three orthogonal directions are 
${\vec p\over |\vec p|}$ and ${\vec {\tilde p}\over |\vec {\tilde p}|}$. 
Since the integrand is an odd
function of $\vec K\cdot \vec {\tilde p}$, the integral will
vanish. Therefore, we conclude that
\be
\Pi_4^{AB} = 0.
\label{b2}\ee

Let us now consider the quantity $\Pi^{\mu\;AB}_{\;\;\mu}$. Using
Eqs. (\ref{4}) and (\ref{16}) we obtain, since $G^\mu_\mu=2$,
\be
\Pi^{\mu\;AB}_{\;\;\mu} = {4g^2N\,\delta^{A,B}\,\over(2\pi)^3}
\int{d^3\vec k\over |\vec k|} N(|\vec k|)
\left[1-\delta^{A,0}\cos(k\cdot\tilde p)\right] .
\label{b3}\ee
\begin{figure}[h!]
\begin{center}
\includegraphics*{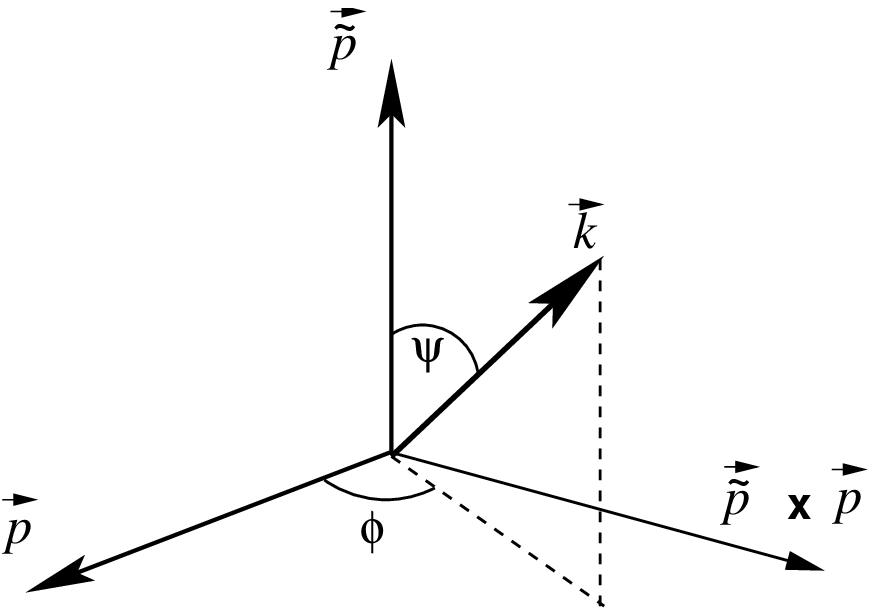}
\end{center}
\caption{}\label{figb1}
  \end{figure}
Using {\it spherical coordinates} as shown in the figure
\ref{figb1}, so that 
\be
k\cdot\tilde p = - |\vec p| \,
\theta\, |\vec k| \, \cos(\psi),
\label{b4}\ee
we can write (with $u\equiv {|\vec k|\over T}$)
\[
\Pi^{\mu\;AB}_{\;\;\mu}  =  {4g^2N\,\delta^{A,B}\,T^2\over(2\pi)^3}
 \int_0^\infty {u du\over {\rm e}^u - 1}
\int_0^{2\pi}d\phi\int_{-1}^{1} d\zeta
\left[1-\delta^{A,0}\cos( \tpt\,u\,\zeta)\right]
\]
\be
= {8g^2N\,\delta^{A,B}\,T^2\over(2\pi)^2} 
 \int_0^\infty {u du\over {\rm e}^u - 1}
\left[1-\delta^{A,0}\;
{\sin( \tpt\,u)\over\tpt\,u}\right] .
\label{b5}\ee
Expressing the Bose distribution in terms of the geometrical series,
\be
{1 \over {\rm e}^u - 1} = \sum_{n=1}^\infty 
{\rm e}^{-n\,u},\;\;\;\;\;
{u \over {\rm e}^u - 1} = - \sum_{n=1}^\infty {d\over dn} 
{\rm e}^{-n\,u}. 
\label{b6}\ee

\be
\int_0^\infty {u du\over {\rm e}^u - 1} =
- \sum_{n=1}^\infty {d\over dn} \int_0^\infty du {\rm e}^{-n\, u} =
 \sum_{n=1}^\infty {1\over n^2} = {\pi^2\over 6}
\label{b7}\ee
and similarly, expanding
\[
\int_0^\infty {\sin(b\, u) du\over {\rm e}^u - 1} =
\sum_{n=1}^\infty\int_0^\infty \sin(b\, u)\, 
{\rm e}^{-n\, u} du
\]
\be=
\sum_{n=1}^\infty \int_0^\infty
{1\over 2i}\left({\rm e}^{(i\,b-n) u} - {\rm e}^{(-i\,b-n) u}\right) du=
b\,\sum_{n=1}^\infty {1\over b^2 + n^2},
\label{b8}\ee
we obtain from (\ref{b5})
\be
\Pi^{\mu\;AB}_{\;\;\mu}  = {8g^2N\,\delta^{A,B}\,T^2\over(2\pi)^2} 
\left[{\pi^2\over 6} -  \delta^{A,0}
\sum_{n=1}^{\infty}{1\over n^2 + \tpt^2}\right]
\label{b9}\ee

\begin{figure}[h!]
\begin{center}
\includegraphics*{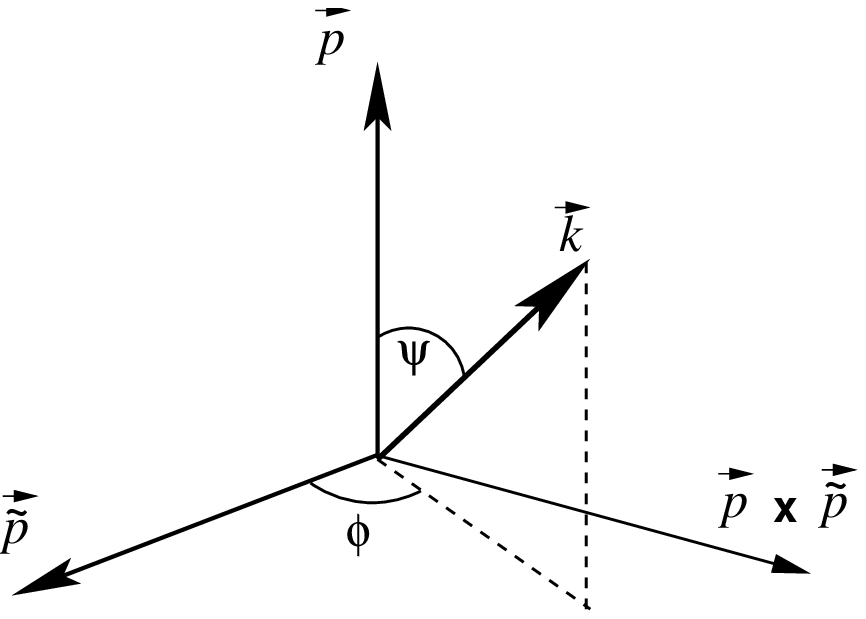}
\end{center}
\caption{}\label{figb2}
  \end{figure}

Let us now consider the component $\Pi_{00}^{AB}$. Using Eqs.
(\ref{4}), (\ref{16}) and (\ref{17})
and the spherical coordinates indicated in figure \ref{figb2}, we obtain 
\begin{eqnarray}
\Pi^{AB}_{00} & = & {2\,g^2N\,\delta^{A,B}\,\,T^2\over(2\pi)^3}
\int_{-1}^{1} d(\cos(\psi))
\int_0^\infty {du\, u \over {\rm e}^u -1}\int_0^{2\pi}d\phi
\left[1 -{2\pop\over \pop - \cos(\psi)} + 
{\pop^2-1\over(\pop - \cos(\psi))^2}\right]
\nnbb &\times &
\left[1-\delta^{A,0}\cos(\tpt\,u\,\sin(\psi)\cos(\phi) )\right].
\;\;\;\;\; 
\label{b10}\end{eqnarray}
Using the standard result \cite{gradshteyn}
\be
\int_0^{2\pi}d\phi\cos(a\,\cos(\phi)) = 2\pi J_0(a),
\label{b11}\ee
we obtain
\begin{eqnarray}
\Pi^{AB}_{00} & = & {2\,g^2N\,\delta^{A,B}\,\,T^2\over(2\pi)^2}
\int_{-1}^{1} d(\cos(\psi))
\int_0^\infty {du\, u \over {\rm e}^u -1} 
\left[1-\delta^{A,0}J_0(\tpt\,u\,\sin(\psi) )\right]
\nnbb &\times &
\left[1 -{2\pop\over \pop - \cos(\psi)} + 
{\pop^2-1\over(\pop - \cos(\psi))^2}\right].
\label{b12}\end{eqnarray}
But now we take 
\be
\int_0^\infty {du\, u \over {\rm e}^u -1} J_0(a\, u) =
-\sum_{n=1}^\infty {d\over dn} \int_0^\infty {\rm e}^{-nu} J_0(a\, u)
\label{b13}\ee
which can be written as \cite{gradshteyn}
\begin{eqnarray}
= 
-\sum_{n=1}^\infty {d\over dn} {1\over \sqrt{n^2+a^2}} =
\sum_{n=1}^\infty {n\over (n^2+a^2)^{3/2}}
& \approx & {\pi^2\over 6}\left(1 - {\pi^2 a^2 \over 10}\right).
\label{b14}\end{eqnarray}
By Eqs. (\ref{b7}) and (\ref{b14}), (\ref{b12}) becomes 
\begin{eqnarray}
\Pi^{AB}_{00} & = & {2\,g^2N\,\delta^{A,B}\,\,T^2\over(2\pi)^2}
\int_{-1}^{1} d\zeta
\left[1 -{2\pop\over \pop - \zeta} + 
{\pop^2-1\over(\pop - \zeta)^2}\right]
\nnbb &\times &
\left[{\pi^2\over 6}-
\delta^{A,0}\,\sum_{n=1}^\infty 
{n\over [n^2+\tpt^2(1-\zeta^2)]^{3/2}}\right] .
\label{b15}\end{eqnarray}

Finally, let us compute 
${\tilde p^\mu \tilde p^\nu\over \tilde p^2} \Pi_{\mu\nu}^{AB}$.
Using Eq. (\ref{4}) and the spherical coordinates indicated in 
figure \ref{figb2}, we obtain 
\begin{eqnarray}
{\tilde p^\mu \tilde p^\nu\over \tilde p^2} \Pi_{\mu\nu}^{AB}
& = & -{2\,g^2N\,\delta^{A,B}\,\,T^2\over(2\pi)^3}
\int_{-1}^{1} d(\cos(\psi))
\int_0^\infty {du\, u \over {\rm e}^u -1}\int_0^{2\pi}d\phi
\nnbb &\times &
\left[{\pop^2-1\over(\pop - \cos(\psi))^2} (\sin(\psi) \cos(\phi))^2-1\right]
\nnbb &\times &
\left[1-\delta^{A,0}\cos(\tpt\,u\,\sin(\psi)\cos(\phi) )\right].
\label{b16}\end{eqnarray}
Performing the integration over $\phi$ \cite{gradshteyn}
\be
\int_0^{2\pi}d\phi\cos^2\phi\,\cos(a\cos(\phi)) = {1\over 2}
\int_0^{2\pi}d\phi(1+\cos(2\phi))\cos(a\cos(\phi)) =
\pi\left[J_0(a)  -  J_2(a)\right],
\label{b17}\ee
we obtain
\begin{eqnarray}
{\tilde p^\mu \tilde p^\nu\over \tilde p^2} \Pi_{\mu\nu}^{AB}
& = & -{2\,g^2N\,\delta^{A,B}\,\,T^2\over(2\pi)^2}
\int_{-1}^{1} d(\cos(\psi))
\left(-\sum_{n=1}^\infty{d\over dn}
\int_0^\infty {\rm e}^{-n\,u}du\right)
\nnbb &\times &
\left\{\left[{\pop^2-1\over(\pop - \cos(\psi))^2} 
{\sin^2\psi\over 2}-1\right]
\left[1-J_0(\tpt\, u\,\sin(\psi))\,\delta^{A,0}\right]
\right.\nnbb & + & \left.
{\pop^2-1\over(\pop - \cos(\psi))^2}{\sin^2\psi\over 2}
J_2(\tpt\, u\,\sin(\psi))\delta^{A,0}\right\}.
\label{b18}\end{eqnarray}
The integral over $u$ in Eq. (\ref{b18}) is first expanded.
Similarly to the Eqs. (\ref{b7}) and (\ref{b13}), we now have
\be
\int_0^\infty {du\, u \over {\rm e}^u -1} J_2(a\, u) =
-\sum_{n=1}^\infty {d\over dn} \int_0^\infty {\rm e}^{-nu} J_2(a\, u)
\label{b19}\ee
which by standard formula in \cite{gradshteyn} becomes
\begin{eqnarray}
-\sum_{n=1}^\infty {d\over dn} \int_0^\infty {\rm e}^{-nu} J_2(a\, u)
& = &  -\sum_{n=1}^\infty {d\over dn} 
{a^{-2} (\sqrt{n^2+a^2} -n)^2 \over \sqrt{n^2+a^2}} \nnbb
& = & {1\over a^2} \sum_{n=1}^\infty 
{(n-\sqrt{n^2+a^2})^2(2\sqrt{n^2+a^2}+n)\over (n^2+a^2)^{3/2}}
\nnbb
& \approx & {3\over 4} a^2 \sum_{n=1}^\infty {1\over n^4}
= {\pi^4 a^2\over 120}.
\label{b20}\end{eqnarray}
One can now reduce 
${\tilde p^\mu \tilde p^\nu\over \tilde p^2} \Pi_{\mu\nu}^{AB}$ to the
expression involving a sum and an integral over $\zeta=\cos(\psi)$ given
in Eq. (\ref{28}).

\end{document}